\documentclass[twocolumn,showpacs,prb]{revtex4}
\usepackage{graphicx}
\usepackage{dcolumn}% Align table columns on decimal point
\usepackage{bm}% bold math

\newcolumntype{.}{D{.}{.}{-1}} % Column type for tables, aligns
\begin{document}

\title{Effect of metal vacancies on the electronic band\\
structure of hexagonal Nb, Zr and Y diborides}
\author{I.R. Shein$^*$, K.I. Shein, N.I. Medvedeva and A.L. Ivanovskii}

\affiliation{Institute of Solid State Chemistry, Ural Branch of
the Russian Academy of Sciences 620219, Ekaterinburg, Russia}

\date{05 November 2002}

\begin{abstract}
Energy band structures of metal-deficient hexagonal diborides
M$_{0.75}$B$_2$ (M = Nb, Zr and Y) were calculated using the
full-potential LMTO method. The metal vacancies change the density
of states near the Fermi level and this effect is quite different
for III-V group transition metal diborides. Contradictory data on
superconductivity in diborides may be supposed to be connected
with nonstoichiometry of samples. Vacancy formation energies are
estimated and analyzed.\\

$^*$ E-mail: shein@ihim.uran.ru
\end{abstract}


% 74.70.-b Superconducting materials (excluding high-Tc compounds)
% 71.20.Dg Alkali and alkaline earth metals
% 71.20.-b Electron density of states and band structure of
%          crystalline solids

\maketitle

A recent discovery of superconductivity (SC) in MgB$_2$ (T$_c$
$\approx$ 39 K) \cite{Akimitsu} and creation of promising
materials based thereon (see reviews$^{[2-5]}$) have attracted a
great deal of interest in hexagonal transition metal (M) diborides
isostructural with MgB$_2$$^{[2-5]}$. Earlier investigations
observed no superconductivity in d-metal diborides (Ti, Zr, Hf, V,
Ta, Cr, Mo, Nb) with temperatures above 0.6 K \cite{Leyarovsky}.
Experimental and theoretical investigations$^{[7-17]}$ showed that
the supercoductivity with T$_c$ $\approx$ 40K is unlikely in all
undoped diborides except MgB$_2$, where the high value of T$_c$ is
explained by strong electron-phonon coupling and high phonon
frequencies and is closely connected with the existence of hole
doped boron $\sigma$ bands.\

Recently, rather high T$_c$ have been reported for ZrB$_2$ (5.5K)
\cite{Gasparov}, TaB$_2$ (9.5K) \cite{Kraczorwski} and NbB2 (5.2K)
\cite{Ogita}. On the other hand, Rosner et al showed
\cite{Rosner1} that SC in TaB$_2$ was absent down to 1.5 K, and
according to \cite{Gasparov,Kraczorwski} the SC transition for
NbB$_2$ is not observed to T $\sim$ 2K, in conformity with earlier
data \cite{Leyarovsky}. The weak electron-phonon coupling
estimated to be inconsistent with superconductivity in ZrB$_2$
\cite{Rosner3} leads to T$_c$ $\sim$ 0.1 K in TaB$_2$
\cite{Rosner2} and is responsible only for T$_c$ $\sim$ 3K in
NbB$_2$\cite{Singh}. So, both experimental and theoretical studies
performed to characterize and understand superconductivity in
MB$_2$ yielded contradictory results. It is important also to note
that in all the studies$^{[7-17]}$ the composition of transition
metal diborides was considered to be strictly stoichiometric (B/M
= 2). As is known, lattice vacancies are typical defects, and
their presence may change considerably the properties of
non-stoichiometric materials in the homogeneity region. For
instance, one of the most familiar classes of non-stoichiometric
compounds with exclusively wide homogeneity regions (to 30 $\sim$
55 at \%) includes cubic (B1) III-V group transition metal
carbides, nitrides and oxides extensively investigated in numerous
works, see \cite{Gusev}. As distinct from the above B1 –-
carbides, nitrides and oxides, III-V group metal diborides have
very narrow homogeneity regions under equilibrium conditions
\cite{Samsonov}. Therefore the role of non-stoichiometry is
usually ignored when their properties are examined. No theoretical
studies describing the effect of lattice vacancies on the
electronic structure of metal diborides are familiar to us
either.\\ Quite recently, Yamamoto et al \cite{Yamamoto} have
carried out a high-pressure synthesis of a series of
metal-deficient samples of the AlB$_2$-type Nb$_{1-x}$B$_2$,
Ta$_{1-x}$B$_2$ (0 $<$ x $<$ 0.48) and Mo$_2$B$_5$-type borides
with the compositions of MoB$_2$ and Mo$_{0.84}$B$_{2.0}$. It was
also found that for Nb$_{1-x}$B$_2$ the superconductivity appears
above x $\sim$ 0.04. As vacancy concentration increases, T$_c$
grows with the maximum value above 9K at x $\sim$ 0.24.\\ In this
report, we are focusing on the effects of metal lattice
non-stoichiometry on the electronic properties of metal diborides.
For this purpose, the band structure of hexagonal NbB$_2$, ZrB$_2$
and YB$_2$ containing 25 \% M-vacancies was investigated
theoretically. The choice of the phases allowed us to discuss the
role of metallic vacancies in modifying the band structure of
isostructural MB$_2$ phases for all possible types of band filling
of these materials.\\ The above diborides have a hexagonal crystal
structure (AlB$_2$-type, space group P6/mmm) composed of layers of
trigonal prisms of M atoms in the center of boron atoms, which
form planar graphite-like networks. The metal-deficient phases
M$_{0.75}$B$_2$ were simulated by 12-atom supercells
(2$\times$2$\times$1). Band structure calculations were carried
out using a scalar relativistic full-potential linear muffin-tin
method (FLMTO) with the generalized gradient approximation (GGA)
for correlation and exchange effects \cite{Method-FLMTO}. Lattice
parameters of NbB$_2$, ZrB$_2$ and YB$_2$ were taken from
\cite{Villars}. As follows from the experiment \cite{Yamamoto},
the lattice constants for the metal-deficient Nb$_{0.75}$B$_2$ (a
= 3.098 \AA,  c/a = 1.072) differ from those for the
stoichiometric NbB$_2$ by less than $\sim$ 0.3 \%. Therefore the
lattice constants for the hypothetical Zr$_{0.75}$B$_2$ and
Y$_{0.75}$B$_2$ were assumed equal to the a, c values of complete
phases \cite{Villars}.\\ Calculated total densities of states
(TDOS) and site-projected l-decomposed DOS (LDOS) of NbB$_2$,
ZrB$_2$ and YB$_2$ and metal-deficient Nb$_{0.75}$B$_2$,
Zr$_{0.75}$B$_2$ and Y$_{0.75}$B$_2$ are shown in Figs. 1 and 2,
respectively. The band structures of AlB$_2$-like 4d-metal
diborides are represented by bonding states formed by B2s and
mixed B2p-Md states (A and B, Fig. 1) separated from the
antibonding states (C) by a pseudogap. Depending on the M
sublattice (the number of valence electrons n$_e$), three types of
band filling of diborides are possible. For ZrB$_2$ (n$_e$ =3.33
e/atom), the Fermi level (E$_F$) is located in the pseudogap (TDOS
minimum). This corresponds to the maximum of chemical stability
when the bonding states are completely occupied and antibonding
states are vacant, see also$^{[23-25]}$. When going to NbB$_2$
(n$_e$ =3.66 e/atom), the antibonding states become partially
occupied and the TDOS at the Fermi level (N(E$_F$)) increases,
Table 1. On the contrary, for YB$_2$ (n$_e$ =3.0 e/atom) some
bonding d-p states are partially unfilled. This determines a
decrease in the cohesive properties of NbB$_2$ and YB$_2$ as
compared with ZrB$_2$. This simple picture is consistent with the
experimental data \cite{Samsonov} and the first-principle
calculations of the formation energies for these diborides,
see$^{[23-25]}$ and below.\\ Let us consider the main effects of M
vacancies on the band structure of these diborides. The
modification of electronic properties when going from MB$_2$ to
MB$_{0.75}$ is generally determined by (i) a decrease in n$_e$ and
(ii) changes of electronic states of atoms surrounding M
vacancies. According to the results obtained, these changes are
quite different for III-V group metal diborides. Nb vacancies in
NbB$_2$ result in the appearance of a new DOS peak and in a
variation of DOS distribution in the region of the Fermi level
(Figs. 2-4). The "vacancy" s-like states are below E$_F$. The
modification of valence states near the Nb vacancy is visually
demonstrated in Fig. 5. It is seen that no new Nb-Nb bonds going
through the vacancy are formed. Deformation of Nb charge density
contours takes place reflecting a growth of the electronic density
along the Nb-Nb bond lines in the vicinity of the defect. For
Nb$_{0.75}$B$_2$, N(E$_F$) decreases insignificantly (by $\sim$
1.9 \%) relative to NbB$_2$.\\ In comparison with NbB$_2$, for
ZrB$_2$ the mentioned changes are much more pronounced, Fig. 2.
The Zr vacancies induce a new sharp DOS peak located in the
pseudogap region. As a result, for metal-deficient Zr$_0.75$B$_2$,
N(E$_F$) increases abruptly (from 0.30 to 1.22 states/(eV cell)).
Quite different trends are observed in the band structure of
Y$_{0.75}$B$_2$: E$_F$ is located in the local TDOS minimum, and
N(E$_F$) reduces more than twice – from 0.900 to 0.404 states/(eV
cell). The changes of LDOS contributions near the Fermi level from
M and B states for III-V group metal diborides in the presence of
M vacancies turned out to be different too, Table I. For
Nb$_{0.75}$B$_2$, N(E$_F$) diminished owing to a decrease in the
contribution from Nb4d states (N$^{Nb}$(E$_{F}$)), whereas the
contribution from boron 2p states (N$^B$(E$_F$)) increases. On the
contrary, a sharp growth of N(E$_F$) for Zr$_{0.75}$B$_2$ is due
to growth of both N$^{Zr}$(E$_F$) and N$^B$(E$_F$), whereas a
decrease in N(E$_F$) for Y$_{0.75}$B$_2$ is caused by simultaneous
reduction of N$^Y$(E$_F$) and N$^B$(E$_F$).\\ Obviously, these
changes in the DOS will affect the properties of
non-stoichiometric diborides, which depend, in particular, on the
near-Fermi level electron density. Our estimations of the
electronic specific heat coefficient ($\gamma$, in the free
electron approximation $\gamma$ =($\pi^2$/3)N(E$_F$)k$^2$$_B$) for
complete MB$_2$ and metal-deficient M$_{0.75}$B$_2$ phases are
listed in Table 2 as an example. It is seen that if for MB$_2$
$\gamma$ changes in the order NbB$_2$ $>$ YB$_2$ $>$ ZrB$_2$, for
non-stoichiometric borides this sequence is as follows:
Zr$_{0.75}$B$_2$ $>$ Nb$_{0.75}$B$_2$ $>$ Y$_{0.75}$B$_2$. The
effect of M vacancies on the value of N(E$_F$) is most noticeable
for ZrB$_2$. Thus, one can suppose that the value of T$_c$ $\sim$
5.5 K observed by Gasparov et al \cite{Gasparov} for ZrB$_2$ may
be due not only to the presence of ZrB$_{12}$ \cite{Shein1} but
also to nonstoichiometry in the Zr sublattice.\\ For
Nb$_{0.75}$B$_2$, no considerable changes are found near E$_F$.
Obviously, the important role in the observed increase of T$_c$
for Nb$_{1-x}$B$_2$ \cite{Yamamoto} will belong to the changes in
phonon frequencies with lowering of crystal stability. On the
contrary, a decrease in N(E$_F$) for the non-stoichiometric
Y$_{1-x}$B$_2$ does not allow us to expect superconductivity in
the homogeneity region of yttrium diboride.\\ Finally we discuss
the variation in chemical stability and cohesive properties of
non-stoichiometric diborides from the numerical estimations of the
cohesive energy (E$_{coh}$) and the heat of formation ($\Delta$H)
of MB$_2$ and M$_{0.75}$B$_2$ phases. The cohesive energies of
MB$_2$ and M$_{0.75}$B$_2$ were calculated as:

\begin{equation}
\begin{array}{c}
\displaystyle E_{coh}^{MB_2} = E_{tot}^{MB_2}- \{E_{at}^M +
2E_{at}^B\}\\
\displaystyle E_{coh}^{M_{0.75}B_2} = E_{tot}^{M_{0.75}B_2}-
\{0.75E_{at}^M +
2E_{at}^B\},\\
\end{array}
\end{equation}

where E$^{M}_{at}$, E$^{B}_{at}$ are the total energies of free M
and boron atoms, and E$^{MB_2}_{tot}$, E$^{M_{0.75}B_2}_{tot}$ are
the total energies (per formula units) of MB$_2$ and
M$_{0.75}$B$_2$, respectively. The heats of formation are defined
as:
\begin{equation}
\begin{array}{c}
\displaystyle \Delta H^{MB_2} = (E_{tot}^{M} + 2E_{tot}^B)-E_{tot}^{MB_2}\\
\displaystyle \Delta H^{M_{0.75}B_2} = (0.75E_{tot}^{M} +
2E_{tot}^B)-E_{tot}^{M_{0.75}B_2},\\
\end{array}
\end{equation}

where E$_{tot}^M$, E$_{tot}^B$ are the total energies of elemental
metals and $\alpha$-boron obtained from FLMTO calculations. In
conformity with experiments \cite{Samsonov}, the MB$_2$ phases
have positive values of formation heat. The value of
$\Delta$H$^{MB_2}$ decreases ($\Delta$H(ZrB$_2$) $>$
$\Delta$H(NbB$_2$) $>$ $\Delta$H(YB$_2$)) being maximum for
ZrB$_2$, Table II. It should be noted that the cohesive energies
(characterizing the atomic decomposition of MB$_2$) exhibit a
different sequence: E$^{NbB_2}_{coh}$ $>$ E$^{ZrB_2}_{coh}$ $>$
E$^{YB_2}_{coh}$. The appearance of a vacancy sharply decreases
the crystal stability ($\Delta$H, Table II). The energy of vacancy
formation in the M sublattice was estimated as:

\begin{equation}
E_{vf} = E_{tot}^{MB_2} - E_{tot}^{M_{0.75}B_2} - 0.25E_{tot}^M,\\
\end{equation}

The results obtained indicate (Table II) that the energy of
formation of M vacancies in NbB$_2$ is lower than that in ZrB$_2$
and YB$_2$. Thus, the M-sublattice nonstoiciometry is more
difficult to achieve in ZrB$_2$, which is the most stable diboride
among those considered above. \\ In conclusion, we have presented
the first results of band structure calculations for
metal-deficient Nb, Zr and Y diborides performed by the FLMTO
method. It was established that vacancies are more likely to
appear in diborides of transition metals of III and V groups. The
effect of M vacancies on the cohesion and electronic properties of
III-V group transition metal diborides was found to be determined
by the band filling. The changes in N(E$_F$) caused by M vacancies
are different: N(E$_F$) may decrease (YB$_2$), increase (ZrB$_2$)
or stay almost constant (NbB$_2$).\\

Acknowledgement.\\

This work was supported by the RFBR, grant 02-03-32971.

%%%%%%%%%%%%%%%%%%%%%%%%%%%% Bibliography %%%%%%%%%%%%%%%%%%%%%%%%%%%

\begin{table}
\caption{Total (N(E$_F$)$^{tot}$) and partial (N(E$_F$)$^l$)
density of states at the Fermi level (states/eV/f.u.) and
electronic specific heat coefficients ($\gamma$,
mJ*mol$^{-1}$*K$^{-2}$) for complete and metai-deficient Nb, Zr
and Y diborides.}

\begin{center}
\begin{tabular}{|c|c|c|c|c|c|c|c|c|c|}
\hline
Phase&N(E$_F$)$^{tot}$&N(E$_F$)$^{Ms}$&N(E$_F$)$^{Mp}$&N(E$_F$)$^{Md}$&N(E$_F$)$^{Mf}$&N(E$_F$)$^{Bs}$&N(E$_F$)$^{Bp}$&$\gamma^{Our}$&$\gamma$\cite{Vajeeston}\\
\hline
NbB$_2$&1.01&0.00&0.01&0.65&0.04&0.01&0.13&2.39&2.42\\
Nb$_{0.75}$B$_2$&0.99&0.01&0.02&0.54&0.02&0.01&0.15&2.34&-\\
\hline
ZrB$_2$&0.30&0.00&0.00&0.17&0.01&0.00&0.04&0.71&0.67\\
Zr$_{0.75}$B$_2$&1.22&0.01&0.03&0.43&0.02&0.02&0.33&2.87&-\\
\hline
YB$_2$&0.90&0.01&0.02&0.36&0.01&0.00&0.14&2.12&2.03\\
Y$_{0.75}$B$_2$&0.41&0.01&0.01&0.12&0.01&0.00&0.08&0.96&-\\
\hline
\end{tabular}
\end{center}
\end{table}

\begin{table}
\caption{Heat of formation $\Delta$H, cohesion energy E$_{coh}$
and vacancy formation energy of vacancy formation E$_{vf}$
(Ry/f.u.) for stoichiometric and metal-deficient Nb, Zr and Y
diborides.}
\begin{center}
\begin{tabular}{|c|c|c|c|c|c|}
\hline
System&E$_{coh}^{MB_2}$&$\Delta$H$^{MB_2}$&E$_{coh}^{M_{0.75}B_2}$&$\Delta$H$^{M_{0.75}B_2}$&E$_{vf}$\\
\hline
Nb - B&1.82&0.27&1.57&0.19&0.08\\
\hline
Zr - B&1.72&0.357&1.46&0.22&0.13\\
\hline
Y - B&1.42&0.26&1.23&0.16&0.10\\
\hline
\end{tabular}
\end{center}
\end{table}

\begin{figure}[btp]
\begin{center}
\leavevmode
\includegraphics[width=5 in]{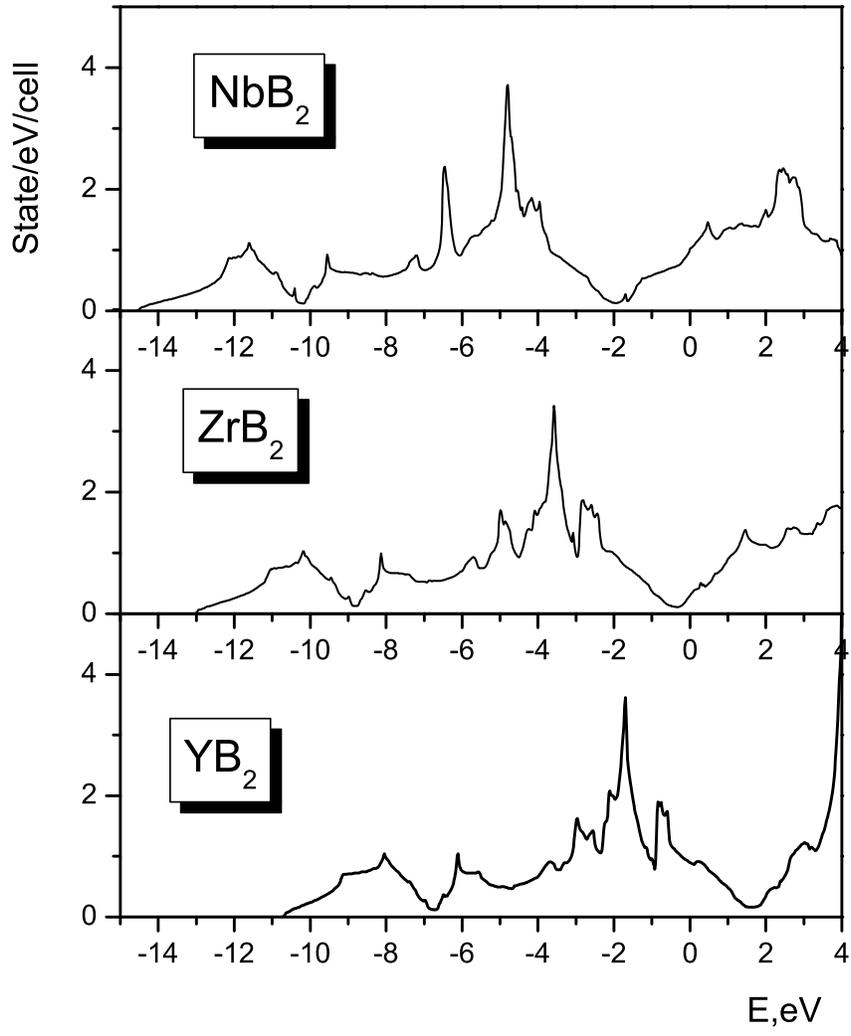}
\caption{The total density of states of NbB$_2$, ZrB$_2$ and
YB$_2$. The energies are relative to the Fermi level.}
\end{center}
\end{figure}

\begin{figure}[btp]
\begin{center}
\leavevmode
\includegraphics[width=5 in]{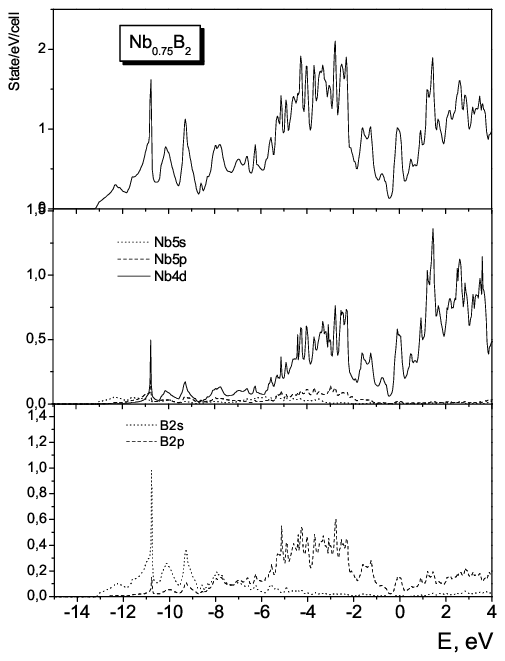}
\caption{The electronic density of states of Nb$_{0.75}$B$_2$.The
energies are relative to the Fermi level.}
\end{center}
\end{figure}

\begin{figure}[btp]
\begin{center}
\leavevmode
\includegraphics[width=5 in]{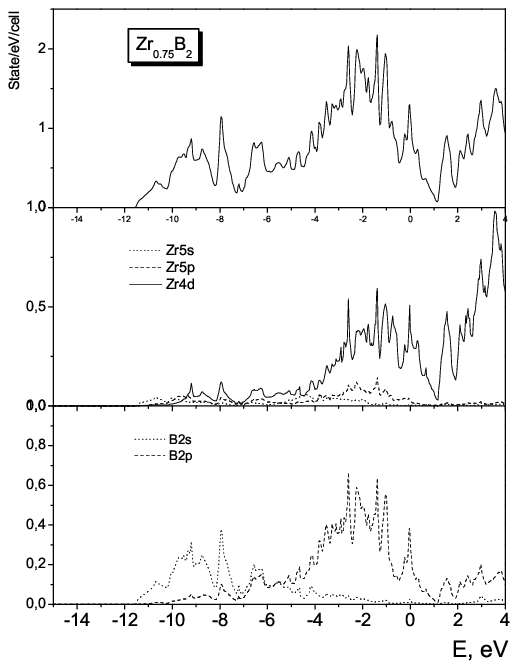}
\caption{The electronic density of states of Zr$_{0.75}$B$_2$. The
energies are relative to the Fermi level.}
\end{center}
\end{figure}

\begin{figure}[btp]
\begin{center}
\leavevmode
\includegraphics[width=5 in]{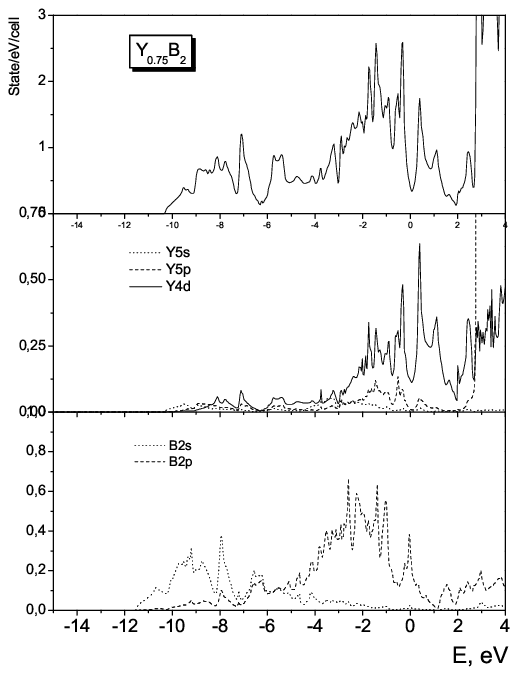}
\caption{The electronic density of states of Y$_{0.75}$B$_2$. The
energies are relative to the Fermi level.}
\end{center}
\end{figure}

\begin{figure}[btp]
\begin{center}
\leavevmode
\includegraphics[width=4 in]{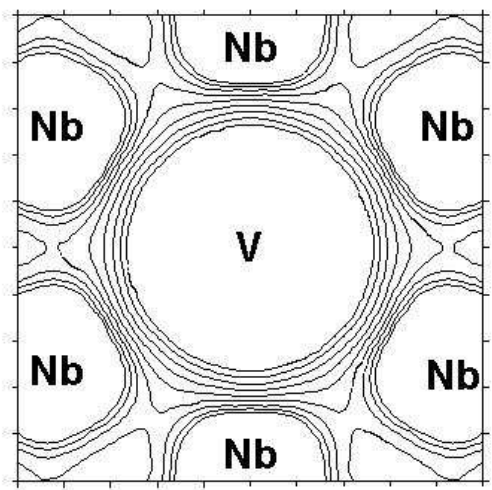}
\caption{Charge density on a hexagonal Nb plane in the vicinity of
metal vacancy (V) in Nb$_{0.75}$B$_2$.}
\end{center}
\end{figure}

\end{document}